\documentclass[a4paper,11pt,dvipsnames,final]{article}
\usepackage{import}
\usepackage{Preamble}

\emailAdd{andrew.d.rolph@gmail.com}
\emailAdd{jackson.r.fliss@gmail.com}

\title{
Unifying energy-entropy bounds in semiclassical gravity 
}
\date{}
\author[a]{Jackson R. Fliss,}
\author[b]{and Andrew Rolph}
\affiliation[a]{Physique Th\'eorique et Math\'ematique, Universit\'e Libre de Bruxelles \& International Solvay Institutes, CP 231, 1050 Bruxelles, Belgique}
\affiliation[b]{Vrije Universiteit Brussel (VUB) and The International Solvay Institutes, Pleinlaan 2, B-1050 Brussels, Belgium}

\abstract{
We revisit bounds on energy and entropy in semiclassical gravity, aiming to strengthen and unify established bounds, in particular the quantum focussing conjecture (QFC) and the smeared null energy condition (SNEC). 
Both the QFC and SNEC have important applications and are the strongest in their respective hierarchies; however, neither implies the other, nor reduces to the null energy condition (NEC) in the classical limit.
To that end, we define a new class of bounds dubbed \textit{gravitational energy-entropy conditions} (GEECs) that imply both the QFC and SNEC, and reduce to the NEC classically. 
We construct an explicit family of GEECs by imposing the defining criteria on a broad set of stress tensor bounds, and constrain its remaining parameter space through test cases, in an effective field theory-inspired approach.
Our GEECs are powerful candidate bounds on energy and entropy in semiclassical gravity.
}

\begin{document}

\maketitle

\section{Introduction}\label{sect:intro}

\begin{figure}
    \centering
    \begin{overpic}[width=0.75\linewidth]{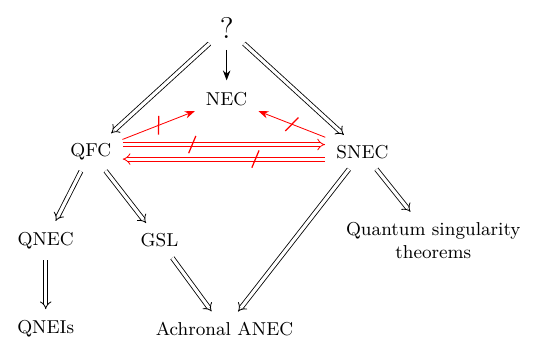}
     \put(8,30){\cite{Bousso:2015mna}}
     \put(24.5,30){\cite{Bousso:2015mna}}
    \put(4,13){\cite{Fliss:2025zhu}}
     \put(36.5,14){\cite{Wall:2009wi}}
     \put(74,31){\cite{Freivogel:2020hiz}}
    \end{overpic}
    \caption{A (non-exhaustive) hierarchy of energy-entropy bounds and their implications.   
    The double-line arrows denote direction of implication. 
    The single-line arrows pointing to NEC denote taking the classical limit. 
    The red arrows emphasise that the QFC and SNEC do not imply each other, nor do they reduce to the NEC in the classical limit.
    Constructing and testing bounds that sit at the top of this hierarchy is a core purpose of this paper. 
    }
    \label{fig:Hierarchy}
\end{figure}

The null energy condition (NEC) is an important assumption in several gravitational theorems, such as Penrose's singularity theorem~\cite{penrose1965gravitational} and Hawking's area theorem~\cite{hawking1972black}.
However, no quantum field theory (QFT) obeys the NEC~\cite{epstein1965nonpositivity}, which is a problem for the application of these gravitational theorems to our universe. 
This motivates finding bounds which are valid for QFTs coupled to gravity and can substitute for the NEC in gravitational theorems, as well as finding
quantum analogues of these theorems, such as the generalised second law (GSL)~\cite{Bekenstein:1974ax}.
Examples of such bounds include the quantum focussing conjecture (QFC), the achronal averaged null energy condition (ANEC), the quantum null energy condition (QNEC), and the quantum Bousso bound.

The broad goal of this paper is to strengthen and unite established semiclassical gravity bounds like these.%
\footnote{By established we mean a well-known bound that either has been proven, at least in a restricted setting, or has accumulated substantial evidence by passing numerous tests.} 
Amongst the set listed above, 
the QFC is special, because it implies all the others~\cite{Bousso:2015mna}. In other words, the QFC sits at the top of a partially ordered hierarchy of logical implications, which includes not only weaker energy-entropy bounds like the GSL and ANEC, but other important implications, such as entanglement wedge nesting~\cite{Akers:2016ugt}. The QFC is also important because several theorems and arguments in the literature assume it, e.g.~\cite{Engelhardt:2014gca, Rolph:2022csa, Headrick:2026ndl}.

This motivates the importance of the QFC, and it is indeed a powerful constraint, but there is one energy bound in semiclassical gravity that is not implied by the QFC: the smeared null energy condition (SNEC), which is the following lower bound on the null-null contracted stress tensor integrated along a segment of a null geodesic against a smearing function $g$:
\bne
    8\pi G_N \int \dd v \, g(v)^2 \langle T_{vv} \rangle_\psi \geq - \int \dd v\, g'(v)^2~. \label{eq:SNEC}\ene
The SNEC, originally proposed in~\cite{Freivogel:2018gxj}, allows for negative null energy density, but limits the amount that light rays can defocus.
The SNEC has important applications, such as in
a generalisation of Penrose's singularity theorem,%
\footnote{The QFC, through the GSL, also implies a singularity theorem~\cite{Wall:2010jtc, Bousso:2025xyc}, but of a different type. More on this in Sec.~\ref{sec:SNEC}.}
and to constrain traversable wormholes~\cite{Freivogel:2020hiz, Bintanja:2023vel}.

The logical independence of the QFC and SNEC bounds, as will be shown in Sec.~\ref{sec:QFC/>SNEC}, suggests that there may be an undiscovered bound that implies them both and so unifies them. 
This is similar to the original motivation for the QFC of uniting the classical focussing theorem and the quantum Bousso bound~\cite{Bousso:2015mna}.
There is another impetus to search for new bounds: neither the SNEC nor the QFC reduce to the NEC in the classical limit,%
\footnote{When the QFC is evaluated on a generic cut of a null geodesic congruence which has either non-vanishing classical expansion or shear. When the expansion and shear do vanish, the QFC reduces to the QNEC, which does become the NEC in the classical limit. More details in Sec.~\ref{sec:QFC}.}
so, in a sense, neither are true quantum generalisations of the NEC. 
While these are hardly unassailable arguments that there has to be a single bound in semiclassical gravity from which all others follow, the possibility is certainly worth investigating, if only because of the principle of parsimony. 

We are aiming to unify the QFC and the SNEC, rather than other bounds, because they sit at the top of their respective hierarchies (as shown in Fig.~\ref{fig:Hierarchy}), so they are the most difficult and constraining bounds to aim for, and unifying them unifies their hierarchies. Unification aside, whether bounds like the QFC and SNEC can be strengthened is of independent interest.

We dub our new bounds \textit{gravitational energy-entropy conditions} (GEECs). GEECs obey three defining criteria: (1) stronger than the SNEC, (2) stronger than the QFC, and (3) reduce to the NEC in the classical limit. The first two criteria mean that GEECs unify the QFC and SNEC, and the third means that they are a true quantum generalisation of the NEC. 

It is non-trivial that there exists a bound that obeys all three GEEC-defining criteria. 
Our goal will be to map out a space of candidate bounds, in an EFT-like spirit, rather than propose a single one.
We will start from an extremely broad set of bounds on the stress tensor, then systematically and drastically reduce the set by imposing the criteria. 
A simple example of one of our new GEECs is
\bne 
    \Theta' \leq - \frac{\theta^2}{D-2} - (\Theta - \theta)^2 - \sigma^2 , \label{eq:L1_intro}
\ene
where $D$ is the spacetime dimension, $\sigma$ the shear, and $\Theta$ and $\theta$ the quantum and classical null expansions, to be defined precisely in the next section. 
It is straightforward to verify that~\eqref{eq:L1_intro} is stronger than the QFC, which states that $\Theta' \leq 0$, since the terms on the right-hand side are non-positive. To verify that it reduces to the NEC in the classical limit, it helps to rewrite~\eqref{eq:L1_intro} as a lower bound on the stress tensor (using Raychaudhuri's equation):
\bne 
    2\pi \langle T_{vv} \rangle_\psi \geq \left( \frac{\hbar S' }{\cA}\right)'+4G_N \left( \frac{\hbar S' }{\cA}\right)^2 ~, \label{eq:Tlow1_intro} 
\ene
where $S'$ is an entropy derivative. Eq.~\eqref{eq:Tlow1_intro} is stronger than the SNEC because it is of the form $8\pi G_N \langle T_{vv} \rangle_\psi \geq X' + X^2 $; we will show that any bound of this form is sufficiently strong to imply the SNEC, and this general form will be our starting point in Sec.~\ref{sec:GEECs} for the construction of a one-parameter family of GEECs we call $\alpha$-GEECs, of which~\eqref{eq:Tlow1_intro} is one member.

Bounds like~\eqref{eq:L1_intro} are powerful candidate constraints on semiclassical gravity. By definition, GEECs imply both the QFC and the SNEC, as well as all bounds and other implications in their two hierarchies, including singularity theorems. Also, some of the $\alpha$-GEECs we will explore in this paper do more than they were constructed for, by naturally implying additional, recently proposed, less established bounds, such as the \textit{strong-QFC} and \textit{field excursion bound} of~\cite{Herderschee:2025nrf}, bringing more bounds under the GEEC umbrella. Also, every member of the $\alpha$-GEEC family implies a quantum analogue of the classical focussing theorem, i.e. initially negative quantum expansions will diverge to $-\infty$ in finite affine parameter.

Since GEECs are so strong, it is especially important to check for counterexamples. The tests that we subject our GEECs to will not rule out all of them, but will reduce the allowed region in parameter space. For example, the simple GEEC given above, \eqref{eq:L1_intro}, is ruled out for $D=4$ CFTs.
We will not prove any of our GEECs. Doing so is at least as difficult as proving the QFC, in the sense that a GEEC proof would automatically be a QFC proof.
Nonetheless, our analysis generates interesting conjectures for future proof attempts. 

The structure of the paper is as follows. We start with background and setup in Sec.~\ref{sec:B&S}, by reviewing classical and semiclassical bounds on the stress tensor, and showing that the QFC and SNEC bounds do not imply each other. 
In Sec.~\ref{sec:GEECs}, we define GEECs and systematically construct an explicit family of them.
To do so, we start
from a general lower bound on the stress tensor that implies the SNEC, then show that, to be stronger than the QFC and to reduce to NEC in the classical limit, uniquely fixes expansion-dependent GEECs up to only a finite number of parameters.
We close with a discussion of our results and future directions in Sec.~\ref{sect:disc}.

\section{Background and setup} \label{sec:B&S}
\subsection{Classical bounds}
The {\it null energy condition} states that the stress tensor double-contracted with any null vector at any spacetime point is non-negative:
\bne T_{vv} \geq 0, \label{eq:NEC} \ene
where $T_{vv}$ denotes
\beq
    T_{vv}:= v^\mu v^\nu T_{\mu\nu}~.
\eeq
The NEC has important implications; for example, it underlies Penrose's singularity theorem~\cite{penrose1965gravitational} and Hawking's area theorem~\cite{hawking1972black}, and it is a sufficient condition to forbid 
traversable wormholes~\cite{Friedman:1993ty}. 

These implications come from how the NEC constrains the dynamics of null geodesics. 
Consider a surface-orthogonal null geodesic congruence in spacetime dimension $D \geq 3$.
Along any generator of the congruence, the expansion $\theta$ obeys the Raychaudhuri equation%
\footnote{There is no rotation (vorticity) term in~\eqref{eq:Ray} because the rotation of surface-orthogonal congruences is zero by Frobenius' theorem.}
\beq\label{eq:Ray}
    \theta'=-\frac{\theta^2}{D-2}-\sigma^2-R_{vv}~,
\eeq
where $\sigma^2 := \sigma_{ab} \sigma^{ab}$ is the shear, and 
the derivative is with respect to the affine parameter $v$ of the generator.
All bounds that follow will transform covariantly under affine reparametrisations $v \to av+b$.

Three equations -- the Raychaudhuri equation, the null-null component of Einstein's equations $R_{vv} = 8\pi G_N\braket{T_{vv}}_\psi$, and the NEC -- together imply the classical focussing theorems
\begin{align}
\label{eq:ClassFocus1}
    &\theta' \leq -\frac{\theta^2}{D-2}~, \\
    \implies &\theta' \leq 0. \label{eq:ClassFocus2}
\end{align}

Both~\eqref{eq:ClassFocus1} and~\eqref{eq:ClassFocus2} state that the expansion never increases along a null geodesic; gravity makes light rays focus. However, there is an important difference between the pair: only~\eqref{eq:ClassFocus1} implies that a caustic will be reached in finite affine parameter if $\theta < 0$ at any point, which is crucial to establish geodesic incompleteness in Penrose's singularity theorem. We will return to this distinction when we discuss strengthening the QFC, which is the quantum analogue of the weaker bound~\eqref{eq:ClassFocus2}.

Most physical classical theories satisfy the NEC, with NEC violation typically implying instability or the existence of superluminal modes~\cite{Dubovsky:2005xd}.
We will assume that the classical limits of our theories obey the NEC.
However, for QFTs, the stress tensor is an unbounded operator~\cite{epstein1965nonpositivity}; in other words, $\inf_{\psi} \langle T_{vv}(x) \rangle_{\psi} = -\infty$. 
So, all QFTs fail to satisfy the NEC, and not just by a small, order $\hbar$ amount, but to an arbitrary degree.
All true, non-trivial, state-independent lower bounds on the stress tensor in a QFT integrate the stress tensor over a spacetime region. Bounds of this type are called \textit{quantum energy inequalities} (QEIs)~\cite{Fewster:2012yh, Kontou:2020bta}. Examples of proven QEIs include the ANEC and the quantum null energy inequalities of~\cite{Fliss:2025zhu}. 
However, many of these proofs are for QFTs, and do not apply in semiclassical gravity. 

\subsection{Bounds in semiclassical gravity} \label{sec:SemiB}
\subsubsection{Semiclassical gravity}
In semiclassical gravity, we treat the gravitational field purely classically, neglecting all graviton loop corrections, including to entanglement entropies.%
\footnote{
We will comment on a graviton loop contribution to entanglement entropy in Sec. \ref{sec:tests} and argue that it does not change the qualitative story presented here. 
}
The semiclassical gravitational effective action is defined from integrating out the matter fields, and is schematically of the form
\bne 
    I_{\rm{s.c.}}(g) =  \frac{1}{16\pi G_N}\int \dd^D x \sqrt{-g} (R + \lambda_1 R^2 + \dots) + W_{\rm{mat.}} (g), \label{eq:highCurv}
\ene
where $W_{\rm{mat.}}$ is the renormalised matter effective action
\bne 
    W_{\rm{mat.}}(g) := -i \log\left ( \int D\phi \, e^{iI_{\rm{mat.}}(\phi,g)}\right) - \text{counter-terms} . \label{eq:Wmat}
\ene
The semiclassical approximation becomes exact when $G_N \hbar \to 0$ with the species scale $N_s G_N\hbar $ held fixed, where $N_s$ is the number of matter species. We assume that the higher curvature corrections in~\eqref{eq:highCurv} are purely induced by matter loops.

Taking the metric variation of $I_{\rm{s.c.}}(g)$ gives the semiclassical Einstein's equations, plus higher curvature corrections:
\bne 
    G_{\mu\nu} + \sum_i \lambda_i (\dots)_{\mu\nu}^{(i)} = 8\pi G_N \langle T_{\mu\nu} \rangle_{\psi},  \label{eq:EinsS}
\ene
where the renormalised stress tensor expectation value is defined as
\footnote{There are many other ways to define the renormalised stress tensor,
such as point-splitting and dimensional regularisation.
There are physical criteria for the renormalised stress tensor that make it unique up to a local geometric conserved tensor (such as $G_{\mu\nu}$)~\cite{Wald:1977up}.
These ambiguities are related to counterterm ambiguities in the action~\eqref{eq:highCurv}.}
\bne 
    \langle T_{\mu\nu} \rangle_{\psi} := -\frac{2}{\sqrt{-g}} \frac{\delta W_{\rm{mat.}}}{\delta g^{\mu\nu}}~.
\ene
The state-dependence of $\langle T_{\mu\nu} \rangle_{\psi}$ comes from the path integral that defines $W_{\rm{mat}}$.

The path integral defining $W_{\rm{mat.}}$ in~\eqref{eq:Wmat} typically has UV divergences which need to be regulated. 
Varying the regulator shifts finite, local, geometric terms between the gravitational and matter parts of $I_{\rm{s.c.}}(g)$. 
This makes the Wilsonian coefficients (including $G_N$) in~\eqref{eq:highCurv} and~\eqref{eq:EinsS} regulator-dependent. 
In what follows, we will neglect higher curvature corrections, but they can be included by replacing $T_{vv}$ in the equations with an effective stress tensor defined by absorbing the higher curvature terms in~\eqref{eq:EinsS} into the stress tensor, or equivalently by replacing $8\pi G_N \langle T_{vv} \rangle_{\psi}$ with (the scheme-independent) $R_{vv}$. 

\subsubsection{The averaged and smeared null energy conditions}~\label{sec:SNEC}

\paragraph{ANEC.} The achronal ANEC states that the stress tensor expectation value integrated along a null line (a complete, achronal null geodesic) is non-negative for any state:
\bne 
    \int_{-\infty}^\infty \dd v \, \langle T_{vv}\rangle_{\psi} \geq 0~.
\ene 
%The ANEC rules out some exotic spacetimes. 

Typical spacetimes do not have null lines,%
\footnote{Spacetimes which obey the null generic condition and certain integrated energy conditions do not have null lines (see Focusing theorem 2 in~\cite{Borde:1987qr}).
}
which limits the applicability of ANEC.
For example, ANEC can rule out short but not long%
\footnote{By long we mean that there exists a null geodesic through the ambient spacetime that can travel from one wormhole endpoint to the other and arrive to the past of any null geodesic passing through the wormhole.}
traversable wormholes
connecting points in the same spacetime, because null geodesics passing through long wormholes are not achronal.
Also, the ANEC is not a useful assumption in putative quantum singularity theorems, because it allows for an arbitrarily large amount of defocussing; the ANEC allows $\int_{v_0}^{v_1} \dd v\langle T_{vv} \rangle_\psi$ to be arbitrarily negative on any compact interval of a null geodesic; it only requires that this be ``paid" for with positive null energy elsewhere.
This motivates the need for energy bounds that are stronger than ANEC and apply to finite geodesic segments.

\paragraph{SNEC.} Freivogel and Krommydas \cite{Freivogel:2018gxj} proposed a generalization of the NEC that is applicable in semiclassical gravity, the {\it smeared null energy condition} (SNEC), which bounds from below the null energy density integrated along a null geodesic $\gamma$:
\beq\label{eq:SNEC2}
    \int\dd v \,g^2\,\langle T_{vv}\rangle_\psi \geq - \frac{1}{8\pi G_N}\int\dd v\,g'^2~,
\eeq
The variable $v$ is an affine parameterisation of the geodesic, $g$ a smearing function, and the null geodesic $\gamma$ we integrate along is achronal on the support of $g$. 

The prefactor on the right-hand side of~\eqref{eq:SNEC} was not specified in the original paper \cite{Freivogel:2018gxj}. 
Later, SNEC was proven for holographic CFTs coupled to dynamical gravity through braneworld holography \cite{Leichenauer:2018tnq} with the prefactor given in \eqref{eq:SNEC}. 
For this paper, when we refer to SNEC, we mean with this ``braneworld prefactor". 

The SNEC can be thought of as a local bound on the stress tensor, since $g$ can have arbitrarily narrow support. The bound trivialises when the width of $g$ limits to zero, as it must do for consistency with the pointwise stress tensor being an unbounded operator. 
The SNEC is stronger than the achronal ANEC, because it implies it when $\gamma$ is a null line, taking $g$ to be constant.

The SNEC is a state-independent lower bound on the smeared stress tensor, in the sense that the right-hand side of~\eqref{eq:SNEC} does not depend on the state $\psi$ either explicitly or implicitly through dependence on the spacetime geometry. In other words, we could drop the $\langle \dots \rangle_{\psi}$ brackets from~\eqref{eq:SNEC} and state it as an operator inequality. The same is true of the ANEC.

Any sufficiently short segment of a null geodesic is achronal, which makes the SNEC more widely applicable than the achronal ANEC. Imposing achronality of $\gamma$ in the SNEC excludes some simple Casimir energy-based counterexamples (which are counterexamples to the non-achronal ANEC too)~\cite{PhysRevD.43.2542}.

The SNEC is important, in part because it is a key assumption in a class of quantum singularity theorems (i.e. singularity theorems with assumptions on the stress tensor that are weaker than the NEC)~\cite{Freivogel:2020hiz}.
There are also quantum singularity theorems that take the GSL as an assumption instead of the SNEC~\cite{Wall:2010jtc}. An advantage of SNEC-based singularity theorems is that the sufficient condition for geodesic incompleteness is purely geometric.%
\footnote{
See~\cite{Freivogel:2020hiz} for a precise statement of the condition.
Essentially, the surface has to be more-than-trapped, because the SNEC allows a (limited) amount of lightray defocussing from negative energy. 
}
In contrast, the condition in the GSL-based theorems is the existence of a so-called quantum trapped surface,%
\footnote{Or in the case of \cite{Bousso:2025xyc}, the validity of an extension of the GSL utilizing the generalized max conditional entropy and the existence of a ``trapped wedge''.} which is a von Neumann entropy-dependent condition and therefore more difficult to verify in practice than a geometric condition.

The evidence that the SNEC is true includes: passing several non-trivial tests~\cite{Freivogel:2018gxj}, a proof in braneworld holography as a consequence of causality on the brane~\cite{Leichenauer:2018tnq}, and an argument for free and super-renormalisable QFTs using ultralocality~\cite{Fliss:2021gdz}.
A general proof of the SNEC has remained elusive; this may be expected since it incorporates directly effects of dynamical gravity.

There is an apparent tension to address between the SNEC and a no-go result by Fewster-Roman which states that there are no universal, non-trivial, state-independent lower bounds on the stress tensor integrated over a finite segment of a null geodesic in QFTs for $D>2$~\cite{Fewster:2002ne}. 
This tension is resolved by the fact that the Fewster-Roman no-go result relies on taking a limit of a particular family of states, and this limit cannot be taken when the QFT has a UV cutoff $\ell_{\tsc{UV}}$~\cite{Freivogel:2018gxj}. In semiclassical gravity, we cannot take $\ell_{\tsc{UV}}$ to be smaller than the Planck scale (or more generally, the species scale), so the no-go result is evaded. 
In a sense, gravity acts as a natural UV regulator.
We could apply the Fewster-Roman argument if we took the limit $\ell_{\tsc{UV}} \to 0$, but we would first have to take the species scale to zero, and the SNEC cannot be violated in this non-gravitational limit, because the inequality trivialises, so everything is consistent.

\subsubsection{The quantum focussing conjecture} \label{sec:QFC}

\begin{figure}
    \centering
    \includegraphics[width=0.5\linewidth]{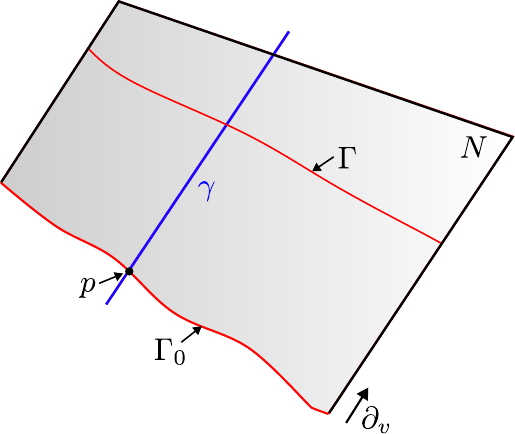}
    \caption{An illustration of a null geodesic $\gamma$ that we integrate the stress tensor along to define the SNEC, a null hypersurface $N$ that contains $\gamma$, and cuts $\Gamma_0$ and $\Gamma$ of $N$.}
    \label{fig:NullHyper}
\end{figure}

Consider a codimension-2 Cauchy-splitting surface $\Gamma_0$ that intersects a null geodesic $\gamma$ at point $p$ and is orthogonal to the tangent vector of $\gamma$ there.
$\Gamma_0$ has two null normals, each with a future- and past-directed orientation. 
We pick the null normal $k^\mu$ on $\Gamma_0$ that is in the same direction as the tangent vector of $\gamma$ at $p$, normalised so that $k^\mu (p)= v^\mu (p)$. 
Following the null geodesics (generators) shot out from each point on $\Gamma_0$ with initial tangent vector $k^\mu$ generates a null hypersurface $N$, of which $\gamma$ is one generator. We terminate $N$ before any generator reaches a caustic or crossing point, so that it is achronal.
See Fig.~\ref{fig:NullHyper} for an illustration.

We use coordinates $(v,y^i)$ on $N$ adapted to $\Gamma_0$: $y^i$ are coordinates on $\Gamma_0$ and each generator is a constant $y^i$. The null geodesic $\gamma$ is at $y^i = y^i_0$. $v$ is the affine parameter along each generator, with $v=0$ the $\Gamma_0$ surface, and the tangent vector $(\del_v)^\mu$ equals $k^\mu$ on $\Gamma_0$. 
Because $\del_v$ is null and normal to $N$, the $g_{vv}$ and $g_{vi}$ metric components are zero, so the induced metric is $h_{ij} = g_{ij}$. The expansion of the congruence is
\beq
    \theta = \del_v \log \sqrt{h}~.
\eeq
For fixed $\gamma$ and $p$, we can always choose a $\Gamma_0$ such that the expansion and shear of the congruence vanish at $p$, but they will then typically be non-zero elsewhere along $\gamma$, because the Ricci tensor acts as a source for expansion, and the Weyl tensor for shear. 

A cut $\Gamma \subset N$ intersects each generator exactly once. 
It is specified by an embedding function $V$:
\bne 
    \Gamma = \{v= V(y)\}~.
\ene
Cuts are Cauchy-splitting, because the subregion of $N$ between $\Gamma_0$ and $\Gamma$ is achronal and $\Gamma_0$ is Cauchy-splitting by assumption.
So, each cut has an associated generalised entropy:
\beq
    S_\gen(V)=\frac{A(V)}{4G_N\hbar}+S(V), \label{eq:Sgen}
\eeq
where $A$ is the area of $\Gamma$ and $S$ is the von Neumann entropy of the state $\psi$ reduced to one side of $\Gamma$.%
\footnote{For a pure state, it does not matter which side, since their von Neumann entropies are the same. For mixed states, the entropies are not the same, so each entropy-dependent bound effectively comes in a pair, depending on which side of the surface one reduces on. The QFC is conjectured to hold for both sides, and we assert the same for our GEECs.
}
In Einstein gravity plus higher curvature corrections, the generalised entropy~\eqref{eq:Sgen} has geometric corrections~\cite{Camps:2013zua, Dong:2013qoa, Harper:2018sdd}, which we neglect here.

The {\it quantum expansion} $\Theta$  in the null direction $\del_v$ at a point $y$ on cut $\Gamma$ is defined as the functional derivative per unit area of the generalised entropy:
\bne 
\begin{split}
    \Theta(V,y) &:= \frac{4G_N\hbar}{\sqrt{h(y)}}\frac{\delta S_\gen(V)}{\delta V(y)}\\
    &= \theta(V(y),y) + \frac{4G_N\hbar}{\sqrt{h(y)}}\frac{\delta S(V)}{\delta V(y)}. \label{eq:ThetaDef}
\end{split} 
\ene
It is the quantum generalisation of the classical expansion $\theta$.

Now we consider a one-parameter family of cuts of $N$ defined by an embedding function 
\bne
    V_v(y):= V_0 (y) + v \dot V (y)
\ene
with $\dot V \geq 0$.
The {\it quantum focussing conjecture} (QFC) takes the $\theta' \leq 0$ relation from~\eqref{eq:ClassFocus2} and posits that it holds in semiclassical gravity if we replace $\theta$ with the quantum expansion~\cite{Bousso:2015mna}:
\bne 
    \frac{\dd\Theta(V_v,y)}{\dd v} \leq 0 . \label{eq:NLQFC}
\ene
This is the non-local form of the QFC. When applied to a cut with vanishing expansion and shear, it gives the QNEC (in non-local form):
\bne 
    \int \dd^{D-2} y \sqrt{h}\, \langle T_{vv} \rangle_{\psi} \dot{V}(y)^2 \geq \frac{\hbar}{2\pi} \frac{\dd^2 S}{\dd v^2} \label{eq:non-local-QNEC}
\ene

The $\dd^2 S/\dd v^2$ term in~\eqref{eq:non-local-QNEC} can be expressed in terms of functional derivatives as follows:
\bne
    \frac{\dd^2 S}{\dd v^2} = \int \dd^{D-2} y \, \dd^{D-2} y' \frac{\delta^2 S(V)}{\delta V(y) \delta V(y')} \dot V (y) \dot V(y').
\ene
The entropy depends non-locally on the cut, so its second-order variation contains both diagonal and off-diagonal pieces. The off-diagonal piece is non-positive, by strong subadditivity~\cite{Bousso:2015mna}. 

The QFC and QNEC reduce to their local forms by choosing $\dot{V}(y)$ to be a top-hat function that equals $1$ in a neighbourhood of $y_0$ with infinitesimal proper area $\cA$, and zero elsewhere.  We will not write $\lim_{\cA \to 0}$ in our formulas, but it should be implicitly understood. 
The quantum expansion can be written as
\bne 
    \Theta = \theta + 4G_N \hbar \frac{S'}{\cA},
\ene 
where we use $(\dots)'$ to denote $\frac{\dd}{\dd v} (\dots)$ with this particular choice for $\dot V$. 
Derivatives of $S$ scale linearly in $\cA$, so terms like $S'/\cA$ are order one as $\cA \to 0$.
In this limit,~\eqref{eq:NLQFC} becomes the local form of the QFC:
\bne 
    \Theta' \leq 0~,
\ene
and~\eqref{eq:non-local-QNEC} becomes the local QNEC:
\bne 
    2\pi \langle T_{vv} \rangle_\psi \geq \frac{\hbar S''}{\cA} ~,
\ene
The QNEC has been proven for quantum field theories (decoupled from gravity) \cite{Bousso:2015wca, Balakrishnan:2017bjg}. 

Some of the QFC's implications, including the GSL, QNEC and ANEC, \textit{have} been proven from first principles in certain settings~\cite{Wall:2011hj, Balakrishnan:2017bjg, Hartman:2016lgu, Faulkner:2016mzt}. 
There is also a recent proof of the QFC in the restricted setting of perturbative quantum gravity on top of a Killing horizon
background~\cite{Chandrasekaran:2026pnc}.
Some putative counterexamples to the QFC have been explored in the literature~\cite{Fu:2017lps, Rolph:2022csa, Franken:2025gwr}. 
The violations in~\cite{Fu:2017lps} are outside of the semiclassical regime, as explained by~\cite{Leichenauer:2017bmc}. 
The violations of the QFC-like bound for JT gravity discussed in~\cite{Franken:2025gwr} do not uplift to $D \geq 3$.

Neither the SNEC nor QFC reduce to the NEC in the classical limit. 
The classical limit of the QFC is $\theta' \leq 0$, or, when written as a lower bound on the stress tensor:
\bne 
    8\pi G_N \langle T_{vv}\rangle_\psi \geq - \frac{\theta^2}{D-2} - \sigma^2 . \label{eq:classQFC}
\ene
This is clearly generically weaker than the NEC.
That said, for any spacetime point, one can always construct a surface-orthogonal null geodesic congruence such that the classical expansion and shear vanish at that point, for which~\eqref{eq:classQFC} \textit{is} the NEC.
Nonetheless, for a generic congruence, the classical component of the QFC is weaker than the NEC by an order $\hbar^0$ amount. 
When either $\theta$ or $\sigma^2 \neq 0$, the QFC is actually quite a weak bound, because the QFC's leading-order classical component~\eqref{eq:classQFC} cannot be saturated by QFTs whose classical limits obey the NEC.

\subsubsection{QFC variants}
\paragraph{SQFC.} Motivated by this, and by the ethos of replacing classical expansions with quantum expansions, we can consider a strengthened version of the QFC called the  strong-QFC (SQFC):
\beq\label{eq:SQFC}
    \Theta'\leq-\frac{\Theta^2}{D-2}~.
\eeq
The SQFC is a natural strengthening of the QFC because it relates to the stronger classical focussing relation~\eqref{eq:ClassFocus1} as the QFC relates to~\eqref{eq:ClassFocus2}. Unlike the QFC, the classical limit of the SQFC is equivalent to the NEC when the shear vanishes. 

The SQFC implies a ``field excursion bound'' on the change of scalar field expectation values along null geodesics~\cite{Herderschee:2025nrf}. Being stronger than the QFC, the SQFC implies at least as many bounds as the QFC; however, we will show that it is still insufficiently strong to imply the SNEC.
Another shortcoming of the SQFC is that it depends on the renormalised $G_N$, and therefore is not renormalisation scheme-independent~\cite{Herderschee:2025nrf}.
Lastly, the SQFC fails to hold in (3+1)d CFTs, as we will show in Sec.~\ref{sec:tests}. 

\paragraph{RQFC.} The restricted-QFC (RQFC) is a weaker version of the QFC, which states that the QFC applies when the quantum expansion vanishes~\cite{Shahbazi-Moghaddam:2022hbw}:
\beq
    \Theta'\leq 0~,\qquad\text{when}\qquad \Theta=0~.
\eeq
The RQFC passes a number of tests, and has been proven in braneworld models and JT gravity~\cite{Shahbazi-Moghaddam:2022hbw, Franken:2025gwr}.
If a counterexample to the QFC were ever found, then it would also be a counterexample to all GEECs, and the RQFC would be a natural replacement bound to unify with the SNEC (if it is not also ruled out).

\subsection{(S)QFC \texorpdfstring{$\nLeftrightarrow$}{<=>} SNEC}\label{sec:QFC/>SNEC}

The SNEC is not stronger than the QFC, because the SNEC does not constrain entropy. To argue it another way, the classical limit of the SNEC allows for NEC violation, whereas the classical limit of the QFC on congruences with $\theta = \sigma^2 = 0$ is the NEC.
We now argue that the reverse statement, that the QFC does not imply the SNEC, is also true.

Let us attempt to derive the SNEC by integrating the Raychaudhuri equation with smearing function $g^2$ along an achronal null geodesic $\gamma$. The result, after integrating the $g^2 \theta'$ term by parts, is
\bne \label{eq:intRay} 
    \int \dd v\,g^2 \langle T_{vv} \rangle_\psi = \frac{1}{8\pi G_N}\int \dd v\left(2gg'\theta - g^2 \frac{\theta^2}{D-2} - g^2 \sigma^2 \right).
\ene

The question is whether the right-hand side of~\eqref{eq:intRay} is lower-bounded as a function of $\theta$, taking the QFC as a constraint.
The answer is no. It would require an upper bound on $\theta$, which neither the QFC nor the SQFC provide, because they are both satisfied by $\Theta = 0$, which relates $\theta$ to $S'$, but does not bound $\theta$.

To argue it another way, consider the QFC rewritten as a lower bound  on the stress tensor:
\bne 
    \text{QFC} \quad \Leftrightarrow \quad 8\pi G_N \langle T_{vv} \rangle_{\psi} \geq -\frac{\theta^2}{D-2} -\sigma^2 + 4G_N \hbar\left( \frac{S'}{\cA}  \right)'. 
\ene
Without further constraints, the variables on the right-hand side are free and do not prevent the stress tensor from being, for example, an arbitrarily negative constant on $\gamma$, thereby violating any bound on the stress tensor integrated along a null generator.

This last statement may appear in tension with the implication chain QFC $\Rightarrow$ GSL $\Rightarrow$ ANEC (see~\cite{Wall:2009wi, Akers:2016ugt}), i.e. that the QFC implies a bound on an integrated stress tensor, but it is not, because those papers impose additional constraints and assumptions. They impose (physically-motivated) boundary conditions on the entropy gradients at the $v \to \pm \infty$ endpoints of the inextendible null geodesic, and
they assume that $\langle T_{vv} \rangle_{\psi}$ and the spacetime metric admit a perturbative expansion in $G_N \hbar$ about a classical solution. In this perturbative regime, the SNEC~\eqref{eq:SNEC} is trivially true. 
This is because the two sides of the SNEC differ by a factor of $G_N$, so a violation of SNEC requires $\sigma_{g}^2 \langle T_{vv} \rangle_{\psi} \lesssim - G_N^{-1}$, where $\sigma_g$ is the width of the smearing function.
In other words, the SNEC is only a tight constraint on energy when either the stress tensor or the metric are not small corrections to a classical background, whereas the arguments behind the QFC $\Rightarrow$ GSL $\Rightarrow$ ANEC implication chain apply in this perturbative regime only.

Accounting for the achronality of $\gamma$ on the support of $g$ does not change our conclusion that QFC $\nRightarrow$ SNEC, because achronality implies an \textit{upper} bound on the integrated stress tensor~\cite{Leichenauer:2018tnq}:
\beq\label{eq:SNEC_upper_bound}
    \int\dd v g^2\,\langle T_{vv}\rangle_\psi \leq  \frac{D-2}{8\pi G_N}\int\dd v\,g'\,^2~.
\eeq
So, the QFC and SNEC do not imply each other, and we have to search for a new kind of bound that unites them and their hierarchies.

\section{Gravitational energy-entropy conditions} \label{sec:GEECs}
\subsection{Definition} \label{sec:GEEC_def}
We define {\it gravitational energy-entropy conditions} (GEECs) by the following three properties:
\begin{itemize}
    \item \textbf{Criterion 1:} Stronger than the SNEC.
    \item \textbf{Criterion 2:} Stronger than the QFC.
    \item \textbf{Criterion 3:} Reduces to the NEC in the classical limit.
\end{itemize}
The first two criteria mean that GEECs unify the QFC and SNEC bounds, and the third that they are a quantum generalisation of the NEC. 

We also do not want our GEECs to be stronger than necessary, in the sense that SNEC should be the strongest bound on the smeared stress tensor derivable from a given GEEC. In other words, it should be mathematically possible to saturate the SNEC while obeying the GEEC.
This is to exclude bounds that are spuriously stronger than SNEC, like the NEC; while the NEC does imply the SNEC, it is impossible to saturate the SNEC while obeying the NEC (except on null lines with a constant smearing function). 

\subsection{Stronger than SNEC} \label{sec:STS}

Let us start with Criterion 1. We claim that any lower bound of the following form on the pointwise stress tensor along an achronal null geodesic is sufficiently strong to imply the SNEC:
\bne 
    \GT \geq  X' + X^2 + Y~, \label{eq:TXYbound} 
\ene
where $X$ and $Y$ are functions of the geodesic's affine parameter, with $Y$ non-negative.
It is a few lines of algebra to confirm this claim. We first weaken~\eqref{eq:TXYbound} by dropping the non-negative $Y$ term, then integrate it against a smooth smearing function $g^2$, 
and lastly integrate the $g^2 X'$ term by parts to get
\bne 
    8\pi G_N \int \dd v g^2 \langle T_{vv} \rangle_\psi \geq -  \int \dd v g'^{2} +  \int \dd v (gX- g')^2 + [g^2 X]^{+\infty}_{-\infty}. 
\ene
From this, we drop the non-negative second term on the right-hand side of this inequality (i.e. weakening it further), and, with the additional (mild) assumption that $\lim_{v\to \pm \infty} (g^2 X) = 0$, the third term too. Then we are done because the result is the SNEC:
\bne 
    8\pi G_N \int \dd v \, g^2 \langle T_{vv} \rangle_\psi \geq - \int \dd v\, g'^{2}~. \label{eq:SNEC3}
\ene
The SNEC is not a quantum bound, by which we mean that~\eqref{eq:SNEC2} has no explicit $\hbar$ dependence. The GEECs that we will construct \textit{are} quantum bounds, and this $\hbar$-dependence is lost when they are weakened to the SNEC in the above derivation. 

The inspiration for~\eqref{eq:TXYbound} comes from~\cite{Fliss:2025zhu}, where we derived something similar to the SNEC (the Fewster-Hollands bound) from the 2d QNEC, which takes a form similar to~\eqref{eq:TXYbound}. 

Note that a non-zero $Y$ is not necessary for~\eqref{eq:TXYbound} to be stronger than SNEC. 
Moreover, $Y = 0$ gives the weakest bound on the stress tensor, which is what we want in general, because then the bound has the best chance of being true. Nonetheless, we keep $Y$ because it will sometimes be needed to satisfy Criterion 3. The classical component of $Y$ will be fixed by Criterion 3, and the quantum component will be constrained by consistency constraints and our tests. $X$ will be tightly constrained by Criterion 2.

\subsection{Stronger than QFC}
We now explore the implications of Criterion 2 for constraining our ansatz \eqref{eq:TXYbound}. To help see when~\eqref{eq:TXYbound} is stronger than the QFC, we rewrite it in QFC-like form (using Raychaudhuri's equation):
\bne 
\text{\eqref{eq:TXYbound}} \quad \Leftrightarrow \quad
    (\theta+X)' \leq - \frac{\theta^2}{D-2} - X^2 - \sigma^2 - Y~. \label{eq:Xgeneq}
\ene
This formula immediately suggests $\Theta - \theta$ as a promising choice for $X$, because \eqref{eq:Xgeneq} is clearly stronger than the QFC for $X=\Theta-\theta$. To constrain $Y$, note that the choice $X = \Theta -\theta$ vanishes as $\hbar \to 0$, so Criterion 3 requires that the classical component of $Y$ is zero to get NEC. 

Making the substitutions $X \mapsto \Theta - \theta$ and $Y \mapsto 0$ (implicitly choosing $Y$ to have no quantum component) in~\eqref{eq:Xgeneq} gives
\bne 
    \Theta' \leq - \frac{\theta^2}{D-2} - (\Theta - \theta)^2 - \sigma^2 . \label{eq:L1}
\ene
This is the GEEC that we gave as an example in Sec.~\ref{sect:intro}. It is indeed stronger than both the QFC and SNEC, and equivalent to the NEC in the classical limit. When expressed in QNEC-like form,~\eqref{eq:L1} becomes~\eqref{eq:Tlow1_intro}. 
While this simple GEEC example illustrates the essential features of the bounds we want to construct, we will see in Sec.~\ref{sec:tests} that it is violated for $(3+1)$d CFTs. Thus we now explore a broad generalisation of $X = \Theta -\theta$: let $X$ be an arbitrary function of the classical and quantum expansions and their derivatives, which we write as
\bne X = f(\Theta, \theta, \Theta', \theta', \dots) - \theta . \label{eq:general_f} \ene
First note that, to satisfy Criterion 2, $f$ cannot depend on the derivatives of $\theta$ or $\Theta$.
For example, if $f$ depended on $\Theta'$, then the $X'$ term in~\eqref{eq:TXYbound} would give a $\Theta''\, \del_{\Theta'} f$ term and~\eqref{eq:TXYbound} would not imply the QFC, because $\Theta''$ is independent of $\Theta'$, takes both signs, and is unbounded in magnitude.

There are further constraints on $f$ from Criterion 2. Plugging the ansatz~\eqref{eq:general_f} into~\eqref{eq:Xgeneq} gives
\bne 
    (\del_\Theta f + \del_\theta f) \Theta' \leq - \frac{\theta^2}{D-2} - (f-\theta)^2 - \sigma^2 - Y + \left(4G_N \hbar \frac{S' }{\cA}\right)' \del_\theta f \label{eq:someA}
\ene
For this to be stronger than the QFC requires $\del_\theta f = 0$, because otherwise the right-hand side would depend linearly on $(S'/\cA)'$, and this is another quantity that is independent of $\Theta'$, takes both signs, and is unbounded in magnitude. We also need $\del_\Theta f > 0$, so that the left-hand side has the correct sign.

Lastly, for~\eqref{eq:TXYbound} to transform covariantly under rescalings of the affine parameter, $f$ must be a homogeneous function of degree one, because the expansions have weight one under affine rescaling. Then, since $\del_\theta f =0$, $f$ must be purely linear in $\Theta$. We parametrise the corresponding $X$ as
\bne 
    X = (\alpha+1) \Theta - \theta , \label{eq:Xgen}
\ene 
with $\alpha > -1$, because $\del_\Theta f > 0$. 

To recap, remarkably,~\eqref{eq:Xgen} is the most general function of the classical and quantum expansions and their derivatives for which~\eqref{eq:Xgeneq} is stronger than the QFC, and transforms covariantly under affine rescaling. 
We chose to consider those $X$ that are functions of the expansions alone, partly because they are a minimal set of ingredients, but also because some other geometric quantities, such as the shear and the other null expansion, cannot be included in $X$ without spoiling Criterion 2, by the same argument given below~\eqref{eq:general_f}. 

\subsection{Reduces to NEC}
Already, the only unfixed quantities left are the parameter $\alpha$ and the non-negative function $Y$.
Now we impose Criterion 3. 
The classical limit of~\eqref{eq:TXYbound} is
\bne 
    (\alpha+1)\GT \geq - \alpha \sigma^2 - \alpha \left(\frac{1}{D-2} - \alpha \right) \theta^2 + Y_{\rm{cl}}. 
\label{eq:class_lim}
\ene
where $Y_{\rm{cl}}$ is the classical limit of $Y$.%
\footnote{Both the stress tensor and geometric quantities like $\theta$ also have $\hbar$ expansions, and, where they appear in~\eqref{eq:class_lim}, they are implicitly the classical components of these quantities. } 
The question is: for what $(\alpha,Y_{\rm{cl}})$ is~\eqref{eq:class_lim} the NEC?
The simplest solution is $(\alpha,Y_{\rm{cl}}) = (0,0)$. More generally,  for~\eqref{eq:class_lim} to be the NEC, i.e. for the right-hand side to vanish, requires
\begin{align}
    Y_{\rm{cl}} &= \alpha \left(
        \sigma^2
        + \left(\frac{1}{D-2} - \alpha\right)\theta^2
    \right), \label{eq:cThetb}
\end{align}
Furthermore, the non-negativity of $Y$ restricts $\alpha$ to the range
\bne 
    \alpha \in \left [0,\frac{1}{D-2} \right]. \label{eq:condition2}
\ene
So, Criterion 3 fully determines the classical component of $Y$, and constrains $\alpha$ to an interval. The quantum component of $Y$, denoted $Y_{q} := Y - Y_{\rm{cl}}$, is still undetermined. 

Note that only for $\alpha = 0$ does the right-hand side of~\eqref{eq:TXYbound} vanish in the classical limit, which one might have thought to be necessary to satisfy Criterion 3. However, this is only sufficient, not necessary; for $\alpha \neq 0$, the right-hand side limits to $-\alpha 8\pi G_N \langle T_{vv} \rangle_{\psi}$, and Criterion 3 is satisfied for $\alpha > -1$.

\subsection{A family of GEECs} \label{sec:AFamily}
We have imposed all the GEEC-defining criteria on the $X$-ansatz~\eqref{eq:general_f} and are ready to write down our explicit bounds. 
The quantum component $Y_q$ was not constrained by the criteria, except through non-negativity of $Y$.
Let us first consider a purely classical $Y$. 
We will see that a non-zero $Y_q$ is important for evading certain counterexamples; however, most of the important features of the GEECs are captured by this simplifying restriction, and we will loosen this restriction in the next subsection.

For purely classical $Y$, only $\alpha$ is unfixed (though constrained to the interval~\eqref{eq:condition2}), and the corresponding one-parameter family of GEECs is
\bne 
    \Theta' \leq -\frac{\theta^2}{D-2} - \sigma^2 - (\Theta-\theta)^2 + \alpha (\theta^2 - \Theta^2), \qquad \alpha \in \left[0,\frac{1}{D-2}\right]. \label{eq:QFCtb} 
\ene 
We call these the \textit{$\alpha$-GEECs}. Let us comment on some of their features.

First we compare the $\alpha$-GEECs to the QNEC. When written as a lower bound on the stress tensor, for easier comparison with the QNEC,~\eqref{eq:QFCtb} becomes%
\footnote{Note that~\eqref{eq:TlowerB} has a $\theta S'$ term (when $\alpha >0$), and yet $\Theta'$ in~\eqref{eq:QFCtb} has a definite sign. This contradicts a claim in~\cite{Shahbazi-Moghaddam:2022hbw} which was made to argue for the existence of QFC counterexamples. }
\bne 
    2\pi \langle  T_{vv}\rangle_\psi \geq \left( \frac{\hbar S' }{\cA}\right)' + 2\alpha \theta \left(  \frac{\hbar S' }{\cA}\right) + 4 (\alpha+1) G_N \left( \frac{\hbar S' }{\cA}\right)^2 , \qquad \alpha \in \left[0,\frac{1}{D-2}\right].
    \label{eq:TlowerB}
\ene
Eq.~\eqref{eq:TlowerB} becomes the QNEC when $S' = 0$, in contrast to the QFC which becomes the QNEC when $\theta =\sigma^2 = 0$. To directly compare~\eqref{eq:TlowerB} to the QNEC, we have to restrict to $\theta = \sigma^2 = 0$, for which~\eqref{eq:TlowerB} becomes
\beq
   2\pi \langle  T_{vv}\rangle_\psi\geq\frac{\hbar S''}{\mc A}+4(\alpha+1)G_N\left(\frac{\hbar S'}{\mc A}\right)^2~,
    \qquad \alpha \in \left[0,\frac{1}{D-2}\right]. \label{eq:gravQNEC}
\eeq
This bound is stronger than the QNEC, with equivalence in the non-gravitational limit $G_N \to 0 $. Note that, while the QNEC is saturated in interacting CFTs with a twist gap \cite{Balakrishnan:2019gxl, Leichenauer:2018obf}, and therefore cannot be strengthened, this does not rule out~\eqref{eq:gravQNEC} since the proofs assume $G_N = 0$.

The ``$\alpha$-QNECs"~\eqref{eq:gravQNEC} strengthen monotonically as $\alpha$ increases. This is different from their $\alpha$-GEEC parents, because whether the right-hand side of~\eqref{eq:QFCtb} increases with increasing $\alpha$ depends on the sign of $(\theta^2 - \Theta^2)$.

The two extremal $\alpha$-values in the range~\eqref{eq:condition2} give interesting special cases of~\eqref{eq:QFCtb}. We have already seen $\alpha = 0$; it gives~\eqref{eq:L1}, whereas $\alpha = \frac{1}{D-2}$ gives
\bne
    \Theta' \leq - \frac{\Theta^2}{D-2} - \sigma^2 - (\Theta-\theta)^2, \label{eq:L2bou}
\ene
or equivalently,
\bne 
    2\pi \langle  T_{vv}\rangle_\psi \geq \left(\frac{\hbar S' }{\cA}\right)' + \frac{2}{D-2} \theta \left( \frac{\hbar S' }{\cA}\right) + 4\frac{D-1}{D-2} G_N \left( \frac{\hbar S' }{\cA}\right)^2~. \label{eq:L2Tbou}
\ene
Eq.~\eqref{eq:L2bou} is interesting because it is similar to, but stronger than, the SQFC~\eqref{eq:SQFC}, and therefore also implies the field excursion bound studied in~\cite{Herderschee:2025nrf}. 

More generally, the $\alpha$-GEECs imply a quantum analogue of the stronger classical focussing theorem~\eqref{eq:ClassFocus1}, derived by maximising the right-hand side of~\eqref{eq:QFCtb} with respect to $\theta$ and dropping the non-positive shear term:

\bne 
    \Theta' \leq - \frac{1+\alpha(1-(D-2)\alpha)}{D-1-(D-2)\alpha}\Theta^2. \label{eq:QFC_real}
\ene
The prefactor on the right-hand side of~\eqref{eq:QFC_real} is negative for $\alpha \in \left[0,\frac{1}{D-2}\right]$, meaning that our $\alpha$-GEECs imply that, if $\Theta$ is negative, then it will diverge to $-\infty$ in finite affine parameter. This is the quantum analogue of a caustic, and has important implications for quantum light-sheets~\cite{Rolph:2022csa, Bousso:2015mna}.

The $\alpha$-GEEC with $\alpha = \frac{1}{2(D-2)}$ is the same as the ``improved" quantum null energy condition (INEC) of~\cite{Ben-Dayan:2023inz, Ben-Dayan:2026bcb}, plus a non-negative gravitational $S'^2$ term which makes it stronger than the INEC. The INEC only holds on surfaces with $\Theta = 0$, whereas our GEECs have no such restriction. 

Our $\alpha$-GEECs, like the SQFC, depend on the renormalised $G_N$ through $\Theta$ (see~\eqref{eq:ThetaDef}), and are therefore scheme-dependent. 
Unlike the QFC, the dependence of $\alpha$-GEECs and the SQFC on $G_N$ cannot be removed, because the difference between their two sides is an inhomogeneous function of $G_N$. While the $\Theta'$ in the QFC also depends on $G_N$, one can divide the QFC by $G_N$ to make it scheme-independent.

This $G_N$-dependence does not mean that the $\alpha$-GEECs are unphysical, only that which renormalised $G_N$ is being used has to be specified. The natural, physical choice is the Newton constant measured at some energy scale $\mu$, which we denote by $G_N^{(\mu)}$, and, of these, $G_N^{\rm IR}$ is the most important from the low-energy EFT viewpoint.
When we come to testing our $\alpha$-GEECs, we will see that some combinations of $\alpha$ and renormalised $G_N$ are ruled out.

The SNEC is scheme-independent, as is manifest when written as a purely geometric inequality:
\bne
    \int \dd v\,(g^2 R_{vv} + g'\,^2) \geq 0~.
\ene
So, a further interesting aspect of the $\alpha$-GEECs is that they are $G_N$-dependent bounds which imply scheme-\textit{independent} bounds (i.e. the QFC and SNEC). 

\subsection{An extended family}
Now we extend our $\alpha$-GEEC family to a larger parameter space by considering non-purely classical $Y$.
The quantum component $Y_q$ can be anything that vanishes in the classical limit and preserves positivity of $Y$, and this $Y_q$ adds simply as a linear quantum correction to the $\alpha$-GEECs (see~\eqref{eq:TXYbound}). 

Consider the most general polynomial function of the classical and quantum expansions that vanishes in the classical limit and transforms with weight two under affine rescalings, which we parametrise as
\bne \frac{Y_q}{1+\alpha} = 2\beta \theta \left(4G_N \hbar \frac{S' }{\cA}\right) + \gamma \left(4G_N \hbar \frac{S' }{\cA}\right)^2, \label{eq:Yq} \ene
with dimensionless parameters $\beta$ and $\gamma$. For the non-negativity of $Y$, we require
\bne \gamma \geq \frac{\alpha+1}{\alpha \left (\frac{1}{D-2} -\alpha \right)} \beta^2. \label{eq:gambo}\ene
This forces $\beta$ to be zero when $\alpha = 0$ or $1/(D-2)$, but it can otherwise take any value in $\mathtt{R}$.
With the quantum component~\eqref{eq:Yq} of $Y$ included, we get the following extended family of GEECs:
\bne 
    2\pi \langle  T_{vv}\rangle_\psi \geq \left( \frac{\hbar S' }{\cA}\right)' + 2\left( \alpha+ \beta\right) \theta \left( \frac{\hbar S' }{\cA}\right) + 4\left(1+ \alpha + \gamma\right) G_N \left( \frac{\hbar S' }{\cA}\right)^2,
    \label{eq:exGEE}
\ene 
or equivalently,
\bne 
    \Theta' \leq -\frac{\theta^2}{D-2} - \sigma^2 + (\alpha+\beta) (\theta^2 - \Theta^2) +(\beta-1-\gamma) (\Theta-\theta)^2
\ene
This extended family reduces to the $\alpha$-GEECs if we set $\beta$ and $\gamma$ to zero. 
Decreasing $\gamma$ weakens the GEEC monotonically, with the weakest allowed value for fixed $\alpha$ and $\beta$ saturating the bound~\eqref{eq:gambo}.
Eq.~\eqref{eq:exGEE} has three parameters for two coefficients, so one parameter can be eliminated, but we choose to keep this redundancy for simplicity of presentation.%
\footnote{Concretely, one can make the substitution $\beta = \delta -\alpha$ in \eqref{eq:exGEE}, set $\gamma$ to its minimal value allowed for by~\eqref{eq:gambo}, and then minimise the resulting $(\hbar S'/\cA)^2$ coefficient with respect to $\alpha \in [0,\frac{1}{D-2}]$. The result is a one-parameter family of GEECs:
\bne 2\pi \langle  T_{vv}\rangle_\psi \geq \left( \frac{\hbar S' }{\cA}\right)' + 2\delta\, \theta \left( \frac{\hbar S' }{\cA}\right) + f(\delta) \,G_N \left( \frac{\hbar S' }{\cA}\right)^2, \ene
with $f$ a messy function, but where the coefficient of the $\theta S'$ term, $2\delta$, can take any value on the real line, unlike the $\alpha$-GEECs.}

The extended family~\eqref{eq:exGEE} is a superset of the $\alpha$-GEECs, so most of the properties discussed in Sec.~\ref{sec:AFamily} carry over. We will see that allowing $Y$ to have a quantum component will be important in certain dimensions in the next subsection. 

\subsection{Tests}\label{sec:tests}
First note that~\eqref{eq:exGEE} cannot be violated by taking the large species number limit, despite being quadratic in the entropy. This is because, while the stress tensor and the entropies both scale linearly with species number $N_{\rm{s}}$, the renormalised Newton's constant scales as $G_N \sim N_{\rm{s}}^{-1}$ as $N_{\rm{s}} \to \infty$~\cite{Kaloper:2015jcz}.

We will primarily test our GEECs with vacuum states in this subsection. We focus on vacuum states because their subregion entanglement entropies are easier to calculate than for generic states, and because they provide stringent tests of GEECs, since they saturate the GEECs' leading-order $\hbar^0$ classical component (i.e. NEC), and so test the next-to-leading-order quantum component. 

All GEECs in the extended family~\eqref{eq:exGEE} are saturated for vacuum states whenever the entropy gradients
$S'$ and $S''$ vanish, such as for flat cuts of a Rindler horizon in the Minkowski vacuum.
Thus we turn to the next simplest setup: spherical entangling surfaces, which will be a sufficiently stringent test to rule out regions of parameter space for~\eqref{eq:exGEE} in certain dimensions.

\subsubsection{(2+1)d vacuum balls}
We start with a (2+1)d CFT in its vacuum state in Minkowski spacetime, for which the generalised entropy of a disk of radius $R$ is
\bne 
    \Sgen = \frac{2\pi R}{4G_N \hbar} + c_0 + c_1 R \label{eq:21Sge}
\ene
where $c_0$ and $c_1$ are constants, with $c_0$ universal and $c_1$ scheme-dependent~\cite{Casini:2011kv}. Shifts in $1/G_N$ can be absorbed by $c_1$, and vice versa. The natural choice of identifying $G_N$ in~\eqref{eq:21Sge} with the physical Newton constant (see discussion in Sec.~\ref{sec:AFamily}) sets $c_1$ to zero.
This is called the area law scheme, because all the area law component of $S_{\rm gen}$ is in the $A/4G_N\hbar$ term. For the area law scheme, the outward quantum expansion for the generalised entropy~\eqref{eq:21Sge} is $\Theta = \theta = R^{-1}$.%
\footnote{For convenience, we have normalised $v$ such that $\dd R/\dd v = 1$. This is not a real restriction, since our bounds transform covariantly under affine rescalings.} 
When $\Theta = \theta$, all members of our extended GEEC family~\eqref{eq:exGEE} reduce to the NEC, so we get no violation.

So, all of the extended GEEC family passes the (2+1)d CFT vacuum ball test. Next we generalise to non-conformal (2+1)d QFTs. We assume that these QFTs have UV and IR fixed-point CFTs, and therefore that the generalised entropy of a vacuum ball asymptotes to~\eqref{eq:21Sge} in the large and small $R$ limits (though the UV and IR coefficients generally differ). 
So, the outward quantum expansion asymptotes to%
\footnote{When taking $R \to 0$, we implicitly hold $R/l_p$ fixed and large, to stay in the semiclassical regime.}
\bne 
    \Theta \sim \left(1+ \frac{2G_N \hbar c_1}{\pi}\right) \frac{1}{R}, \qquad \text{ as } R \to 0,\, \infty. \label{eq:3dquE}
\ene
with $c_1 \to c_1^{\tsc{UV}}$ as $R\to0$, and $c_1 \to c_1^{\tsc{IR}}$ as $R \to\infty$.

What makes the QFT different from the CFT is that $c_1$ cannot be zero in both the UV and the IR. This is because the entanglement entropy of a disk is a concave function of its radius~\cite{Casini:2012ei}, i.e. $\del_R^2 S(R)\leq 0$, due to strong subadditivity. For generic QFTs whose $S(R)$ is non-linear away from the small and large $R$ asymptotics, this implies that
\beq
    c_1^{\tsc{UV}} - c_1^{\tsc{IR}} = - \int_0^\infty \dd R \, \del_R^2 S(R) > 0~.
\label{eq:DeltaC}
\eeq
Plugging in $\langle T_{vv} \rangle_{\psi} = 0$ and~\eqref{eq:3dquE} for $\Theta$ into our extended GEEC family~\eqref{eq:exGEE}, we find that the following region of parameter space is ruled out:
\bne \epsilon\left(\alpha - \frac{1-2\beta-(1+\gamma)\epsilon}{2+\epsilon}\right) > 0 , \qquad \qquad \epsilon := \frac{2 G_N \hbar c_1}{\pi}~.\label{eq:alpeb}\ene
Let us apply this to the $\alpha$-GEECs, i.e.~\eqref{eq:alpeb} with $\beta = \gamma = 0$.
If we choose to use $G_N^{\rm IR}$ in defining the $\alpha$-GEECs, i.e. the natural IR area law scheme, with $c_1^{\rm{IR}} = 0$, then $\epsilon^{\rm{UV}} > 0$
and the $\alpha \gtrsim 1/2$ half of the family are violated in the UV.%
\footnote{We assume that $c_1^{\tsc{UV}} - c_1^{\tsc{IR}}$ is order $(G_N \hbar)^0$, and therefore that $\epsilon$ is perturbatively small.}
Vice versa, we cannot use (the slightly less natural) $G_N^{\rm UV}$ for the $\alpha \lesssim 1/2$ half of the family, because they would be violated in the IR. The $\alpha$-GEEC with $\alpha = 1/2$ is fully ruled out for 2+1d QFTs, because~\eqref{eq:alpeb} with $\alpha = 1/2$ is only avoided if $\epsilon=0$, so this GEEC will unavoidably be violated in either the UV or IR. 
In short, the $\alpha$-GEECs using $G_N^{\rm{IR}}$ for $\alpha \lesssim 1/2$ and $G_N^{\rm UV}$ for $\alpha \gtrsim 1/2$ pass the (2+1)d QFT vacuum ball test.

\subsubsection{(3+1)d vacuum balls} \label{sec:3+1d_balls}
The generalised entropy of a ball-shaped region in Minkowski spacetime for the vacuum state of a (3+1)d CFT is~\cite{Solodukhin:2008dh, Casini:2011kv}
\beq
    \Sgen(R)=\frac{A}{4G_N\hbar}-2a\log A+c_0+c_2 A~, \label{eq:4dbal}
\eeq
where $A = 4\pi R^2$ is the ball's surface area.
The $c_0$ and $c_2$ constants are scheme-dependent, whereas $a$ is the scheme-independent $a$-anomaly coefficient of the CFT. 
From~\eqref{eq:4dbal}, the outward quantum expansion on the ball's boundary is
\beq
    \Theta(R)=\left(1+4G_N\hbar c_2-\frac{2G_N\hbar\,a}{\pi R^2}\right)\frac{2}{R}~.
    \label{eq:theR4}
\eeq
In contrast to (2+1)d, we cannot set the quantum component of this $\Theta$ to zero with a suitable choice of scheme, so we will get a non-trivial test of our GEECs, even for CFTs.

One can show that the quantum expansion~\eqref{eq:theR4} with the area law scheme violates the SQFC~\eqref{eq:SQFC}. The QFC is also violated, but not in the same way: only for ball radii at or below the species scale, i.e. outside of semiclassical gravity's regime of validity.

For the $\alpha$-GEECs, plugging~\eqref{eq:theR4} into~\eqref{eq:TlowerB} gives
\bne 
    0 \stackrel{?}{\geq} \frac{4G_N\hbar}{R^2}\left[(8\alpha-2)c_2+(3-4\alpha)\frac{a}{\pi R^2}\right] + (1+\alpha)\frac{\left(4G_N\hbar\right)^2}{R^2}\left(2c_2-\frac{a}{\pi R^2}\right)^2~. 
\label{eq:4dcou}\ene
Looking at the leading order term in $G_N\hbar$, we see that there is no $\alpha\in[0,1/2]$ which satisfies~\eqref{eq:4dcou} when $a >0$, i.e. any non-trivial unitary CFT$_4$. This is easy to see for the area law scheme, where $c_2 = 0$; however, it is true in any scheme if $a > 0$: \eqref{eq:4dcou} will fail for small enough radius.

Including the graviton contribution to the ball's entanglement entropy does not help the $\alpha$-GEECs evade violation. The violation effectively comes from the negative prefactor of the $\log$ term in~\eqref{eq:4dbal}, and the graviton contribution only exacerbates that negativity~\cite{Benedetti:2019uej}.
Including higher-curvature corrections in the gravitational effective action does not help either, because the corresponding correction to the sphere's generalised entropy~\eqref{eq:4dbal} is an $R$-independent constant.

The upshot is that the CFT vacuum ball is a counterexample for the whole $\alpha$-GEEC family for $D=4$. But large regions of parameter space are still safe for the extended family~\eqref{eq:exGEE}, as we will see next.

\subsubsection{Higher dimensions} In general, even spacetime dimensions, the universal piece of the CFT vacuum ball entropy is~\cite{Casini:2011kv}
\bne S_{\rm univ} = (-1)^{\frac{D}{2}-1} 4 a_D \log R. \label{eq:univ_S}\ene
This alternating sign is inherited by the terms linear in $S'$ in~\eqref{eq:exGEE}.%
\footnote{This assumes that $a_D \geq 0$, which is true for holographic CFTs, but has not been proven for non-holographic CFTs (except $D=4$~\cite{Hofman:2008ar}).}
The consequence is that all the $\alpha$-GEECs pass the $D$-dim CFT vacuum ball test when $D = 2 \pmod 4$, but fail at order $G_N \hbar$ when $D = 0 \pmod 4$, when we work in a generalised area law scheme of setting all the scheme-dependent $c_n$ to zero.

For the extended GEEC family, we evade a CFT vacuum ball counterexample for even $D$ if
\begin{equation}\label{eq:betabound1}
    2(\alpha+\beta) > \frac{D-1}{D-2}\qquad  \text{ for } D = 0 \pmod{4},
\end{equation}
and
\begin{equation}\label{eq:betabound2}
    2(\alpha+\beta) < \frac{D-1}{D-2}\qquad \text{ for } D = 2 \pmod{4}.
\end{equation}
These are strict inequalities because the $S'^2$ term in~\eqref{eq:exGEE} has a positive coefficient.
Note that the vacuum ball does not rule out a purely classical $Y$ in $D=2 \pmod{4}$ dimensions.

CFT vacuum balls in odd spacetime dimensions $D \geq 3$ do not violate any member of the extended GEEC family~\eqref{eq:exGEE}, if we use the generalised area law scheme, for the same reason given after~\eqref{eq:21Sge}: the universal piece is $R$-independent, and so $\Theta = \theta$. For QFT vacuum balls, we have
\bne \Theta \sim \left(1+\frac{2G_N \hbar c_{D-2}^{\tsc{UV}}}{\pi}\right) \frac{D-2}{R}, \qquad \text{as } R \to 0, \label{eq:genDthe}\ene
and we have $c_{D-2}^{\tsc{UV}} \geq c_{D-2}^{\tsc{IR}}$~\cite{Casini:2017vbe}. Eq.~\eqref{eq:genDthe} is the same as~\eqref{eq:3dquE}, except for the replacement $c_1 \mapsto c_{D-2}$, so the resulting constraint on the parameter space is essentially the same in any odd dimension.
This last statement applies when using the IR area law scheme only: while the UV area law scheme implies $c_{D-2}^{\tsc{IR}} \leq 0$, the large $R$ behaviour of $\Theta$ is dominated by $c^{\tsc{IR}}_1 R^{-1}$, and this entropy coefficient is not known to obey irreversibility theorems, except for $D =3$ where $c_{D-2} = c_1$.

\subsection{Comments} \label{sec:GEECcomments}
\paragraph{Other tests.} The vacuum balls we have considered give non-trivial constraints on the $\alpha$-GEECs, but they are still quite a limited family of tests. With additional tests, we could exclude further regions of the parameter space in the extended family of GEECs in~\eqref{eq:exGEE}. 
The obstacle is the difficulty in evaluating entanglement entropy for non-vacuum states and all but the most symmetric spatial regions.
Setups with calculable entanglement entropy, that can therefore provide tractable tests of GEECs, include:
\begin{itemize}
    \item Weakly excited states, because we can use the first law of entanglement, $\Delta S = \Delta K$, in cases where the modular Hamiltonian $K$ is known~\cite{Blanco:2013joa, Rolph:2021nan}.
    \item Braneworld setups, because the entanglement can be calculated holographically.
    \item Wiggly cuts of null cones and null planes~\cite{Casini:2017roe, Casini:2018kzx}. Our vacuum ball tests were effectively non-wiggly cuts of the null cone.
    \item Deformed spheres~\cite{Mezei:2014zla, Allais:2014ata, Faulkner:2015csl}.
\end{itemize}

\paragraph{SNEC saturation.}
It would be useful to have an explicit state that saturates the SNEC, but we do not know of one.
An obstacle to constructing such states is that gravitational backreaction cannot be neglected, because saturating the SNEC requires  $\sigma_g^2\langle T_{vv} \rangle_\psi \approx G_N^{-1} $, where $\sigma_g$ is the width of $g$.
However, we can ask whether our GEECs are too strong to allow the SNEC to be saturated (recall the ``fourth" condition discussed in Sec.~\ref{sec:GEEC_def}).
From the derivation in Sec.~\ref{sec:STS}, for the SNEC to be exactly saturated (for fixed $g$) requires the following on the support of $g$: (1) saturation of the pointwise bound~\eqref{eq:TXYbound}, (2) $Y = 0$, and (3) $X = g'/g$. 
Condition (2) points to a further benefit of including quantum corrections to $Y$: for $Y_\text{cl}$ to vanish for $\alpha \in (0,\frac{1}{D-2})$ requires $\theta = 0$ on the support of $g$, which is too restrictive as it is equivalent to $\langle T_{vv} \rangle_{\psi}$ vanishing. 
Condition (3) implies that $X$ must diverge at the boundary of the smearing region, since $g'/g$ does so when $g$ is compact.
The conditions (1)-(3) have some slack if we only require approximate saturability of SNEC, such as with a non-singular $X$ that only equals $g'/g$ away from the boundaries of $\supp(g)$. Approximate saturability is natural from the perspective of the braneworld proof, whose derived SNEC bound has corrections at finite brane depth~\cite{Leichenauer:2018tnq}.
For the extended GEEC family~\eqref{eq:exGEE} with $\beta \neq0$ and the minimal allowed value for $\gamma$, one can show that the three conditions can be simultaneously satisfied.

\paragraph{Other GEECs.} It would be interesting to construct GEECs built from different ingredients. In this paper, we only considered those $X$ which are functions of $\Theta$, $\theta$ and their derivatives (see~\eqref{eq:general_f}), but one could generalise to other quantities, geometric or otherwise. There are some natural candidates. 
First, since the vacuum state ruled out some of our $\alpha$-GEECs, one could consider replacing entropies with their vacuum-subtracted counterparts,
\bne S \mapsto S - S_{\rm{vac}}, \ene
and thereby trivially evade vacuum ball counterexamples. 
However, the difference in entropies between two states in a given region is only well-defined in a fixed background geometry, so the difference between the backreacted geometries for the vacuum state and $\psi$ must be negligible for $S - S_{\rm{vac}}$ to be well-defined. 
Another issue is the lack of a unique vacuum in curved spacetimes. While it is natural to consider states with similar singularity structures, i.e. Hadamard states, the resulting bound would still be dependent on the choice of Hadamard state. This reflects a similar dependence in regulating the stress tensor one-point function through state subtraction; in our approach to defining the renormalised stress tensor, this ambiguity is reflected in counterterms in \eqref{eq:Wmat}.

It is also natural to consider anomaly subtraction from the entropy, since anomaly terms appear in conformal transformations of the QNEC~\cite{Wall:2011kb, Koeller:2015qmn, Akers:2017ttv}. A problem with anomaly subtraction, and vacuum subtraction too, is that they spoil Criterion 2, for the same reason given after~\eqref{eq:general_f}. 

\section{Discussion}\label{sect:disc}
The broad goal of this paper has been to unite bounds on energy and entropy in semiclassical gravity. 
Within that broad goal, we have focussed on GEECs, defined in Sec.~\ref{sec:GEEC_def}, which unite the QFC, SNEC, and NEC bounds specifically.
From the abstract set of all GEECs, we have constructed and focussed on explicit examples, the $\alpha$-GEECs and the extended family~\eqref{eq:exGEE}.

Uniting and strengthening bounds is important for several reasons. First, the principle of parsimony: why have two logically independent bounds like the QFC and SNEC if one might suffice? Second is the tightness of bounds: GEECs reduce to the NEC in the classical limit for any congruence, whereas the classical limit of the QFC has slack for congruences that are not locally stationary.  
Third is that strong bounds, even if only conjectural, have implications which are interesting in their own right, such as the QNEC from the QFC, the field excursion bound from the strong QFC, and our quantum focussing inequality~\eqref{eq:QFC_real} from $\alpha$-GEECs.

The $\alpha$-GEECs are not a single, unique proposal for a new energy-entropy bound; they are a continuous family. 
This is reflective of our approach to identifying GEECs, which is EFT/bootstrap-like in the sense that we started from~\eqref{eq:TXYbound}, a very general pointwise lower bound on the stress tensor that implies the SNEC, then used the other GEEC-defining criteria and tests to exclude regions of parameter space.%
\footnote{It is not literally a bootstrap approach, because the GEEC criteria are definitions, not consistency conditions or fundamental physical properties.}
An intriguing possibility is if additional tests or physical arguments could shrink the set of allowed GEECs to a set which is totally ordered by logical implication and/or for which there is a strongest GEEC.

The $\alpha$-GEECs and the extended family are a prototypical subset of the set of all GEECs. While we hope that at least one of them is true, they are falsifiable, and it is possible that there is a test we have not considered that rules out the whole family. If any counterexample to the QFC or the SNEC were found, it would also be a counterexample to all GEECs.
Neither of these possibilities would be fatal to our broader programme of strengthening and unifying bounds in quantum gravity; rather, we would see it as additional impetus to broaden the search. 

\subsection{Future directions}

\paragraph{Broadening the GEEC search.} The clearest future direction is to find more GEECs and test them. As shown in Sec.~\ref{sec:STS}, any bound of the form~\eqref{eq:TXYbound} satisfies Criterion 1. We explored those $X$ in~\eqref{eq:TXYbound} which are functions of $\Theta$ and $\theta$ alone, and we showed that the other two criteria and our tests tightly constrain that function, but the broad generality of~\eqref{eq:TXYbound} impels us to explore other possibilities for $X$. We discussed vacuum and anomaly subtraction possibilities in Sec.~\ref{sec:GEECcomments}, and we will discuss scheme-independent possibilities next.

\paragraph{Scheme-independence.} We encountered several subtleties arising from the scheme-dependence of the Wilsonian coefficients in the gravitational effective action, in particular, needing to specify which renormalised $G_N$ to use in the $\alpha$-GEECs and the extended family~\eqref{eq:exGEE}. These subtleties are not fatal, but they do invite us to search for scheme-independent GEECs. Natural scheme-independent quantities to use in $X$ are relative and generalised entropies.
The main obstacle we found to constructing a scheme-independent GEEC was Criterion 2, because the $G_N$-dependence of $\Theta$ drops out of the QFC but not the strengthenings of QFC we explored.

\paragraph{Applications to QFTs.} 
We motivated GEECs for their utility in constraining the geometric side of semiclassical gravity: by design, GEECs imply the SNEC, which implies a semiclassical generalization of the Penrose singularity theorem. 
However, GEECs may also imply extra constraints on which QFTs can be consistently coupled to semiclassical gravity. 
For example, integrating~\eqref{eq:TXYbound} along a null line gives a stronger form of the ANEC:%
\footnote{This assumes that the boundary term $[X]^{\infty}_{-\infty}$ can be dropped, and that for the null geodesic to be complete and achronal does not place additional constraints on $X$ and $Y$.}
\beq
    8\pi G_N\int \dd v \,\langle T_{vv}\rangle_{\psi} \geq \int \dd v\,(X^2+Y)~,
\eeq
which may then imply $G_N$ corrections to the Hofman-Maldacena bounds for 4d CFTs \cite{Hofman:2008ar}. 

\paragraph{Unifying other semiclassical gravity bounds.} The established semiclassical gravity bounds that we chose to attempt to unify in this paper were the QFC and SNEC. These are natural targets because they sit at the top of their respective hierarchies (with neither implying the other), so they are the most difficult and constraining bounds to aim for, and unifying them unifies their whole hierarchies. 
Nonetheless, one could aim to strengthen and/or unify some other set of bounds.
For example, one could aim for bounds weaker than the QFC, like the QNEC or RQFC, as mentioned above. If there are any other established semiclassical gravity bounds that are in neither the QFC nor SNEC hierarchies (we are not aware of any), then these would also be natural targets for unification.

\paragraph{Tube SNEC.} In the case that a counterexample to the SNEC were found, one could aim for a codimension-one version of SNEC modified with an additional smearing envelope over the $D-2$ transverse directions, what we might call a ``tube SNEC." A version of such a bound (in $D> 3$ dimensions) is already implied by integrating the Raychaudhuri equation directly and completing the square, as in Sec. \ref{sec:STS}:\footnote{In $D=3$ there is no square to complete and integrating Raychaudhuri simply does not imply any version of tube SNEC.}
\beq\label{eq:RayTubeSNEC}
    8\pi G_N\int\dd v\int \dd^{D-2}y\sqrt{h}G(v,y)^2\langle T_{vv}(v,y)\rangle_\psi\geq-\frac{D-2}{D-3}\int \dd v\int \dd^{D-2}y\sqrt{h}\left(\pa_vG(v,y)\right)^2~.
\eeq
However, by ``tube SNEC'' we mean the strengthening of this to the braneworld coefficient inspired by the original SNEC:
\beq\label{eq:BWTubeSNEC}
    8\pi G_N\int\dd v\int \dd^{D-2}y\sqrt{h}\,G(v,y)^2\langle T_{vv}(v,y)\rangle_\psi\geq-\int\dd v\int \dd^{D-2}y\sqrt{h}
    \left(\pa_vG(v,y)\right)^2\,.
\eeq
This differs from the SNEC generically, which we see by taking $G(v,y)^2=g(v)^2\,\frac{\delta^{D-2}(y-y_0)}{\sqrt{h(y)}}$:
\beq
    8\pi G_N\int \dd v\,g(v)^2\langle T_{vv}(v,y_0)\rangle_\psi\geq-\int\dd v\left(g'(v)-\frac{1}{2}g(v)\theta(v,y_0)\right)^2~.
\eeq
The conjectural tube SNEC, \eqref{eq:BWTubeSNEC}, is, to our knowledge, also logically distinct from the QFC hierarchy. We find it interesting that the bound \eqref{eq:RayTubeSNEC} is trivial -- beyond the input of Einstein's equations, it follows from a geometric identity, i.e. Raychaudhuri's equation -- yet the tube SNEC with the slightly stronger braneworld coefficient, \eqref{eq:BWTubeSNEC}, seems to be independent of both SNEC and QFC.

\paragraph{The strongest GEEC.} It is desirable to find the strongest (true) bound on the stress tensor in semiclassical gravity, or, better yet, an equality constraint. We claim that the SNEC itself suggests that such an equality might exist. 
The SNEC is equivalent to $\hat H:= \GT - \del_v^2$ being a non-negative operator on the space of smearing functions: 
\beq
    \int \dd v\, g\,\hat H\,g \geq 0~. \label{eq:posH}
\eeq
This formula is mathematically similar to the local minimality of an action. 
To see this, consider an action $I = \int dt (\dot{x}^2/2 - V(x))$, and note the similarity between~\eqref{eq:posH} and the non-negativity of this action's second-order variation:
\bne \int \dd t\, \delta x \left(-V''(x) - \del_t^2 \right) \delta x\geq 0. \ene
With this connection between the SNEC and the calculus of variations identified, we can apply the standard Jacobi condition (for a reminder, see~\cite{Gelfand:1963}) to show that~\eqref{eq:posH} is equivalent to $\hat H \zeta = 0$ being a disconjugate differential equation, which is equivalent to $\hat H \zeta = 0$ having a strictly positive solution ($\zeta>0$)~\cite{derr2008theorydisconjugacysecondorder}.
After the substitution $Z = \zeta'/\zeta$, the equation $\hat H \zeta = 0$ becomes
\bne \GT = Z' + Z^2, \label{eq:Jacobi}\ene
and the existence of a strictly positive $\zeta$-solution becomes the existence of a globally regular $Z$-solution to~\eqref{eq:Jacobi}. 
Global regularity comes from the strict positivity of the $\zeta$-solution; while there are always local regular solutions to equations like~\eqref{eq:Jacobi} for any $\langle T_{vv} \rangle_{\psi}$, \textit{global} regularity is a non-trivial constraint on $\langle T_{vv} \rangle_{\psi}$.

So, the SNEC is equivalent to~\eqref{eq:Jacobi} having a globally regular solution.
Note the similarity between~\eqref{eq:Jacobi} and~\eqref{eq:TXYbound}.
The Jacobi condition does not determine $Z$ for us.
However, supposing that $Z$ could be identified with $\Theta -\theta = 4 G_N \hbar S'/\cA$,
then~\eqref{eq:Jacobi} is identical to QNEC saturation (which has been proven for interacting CFTs with a twist gap~\cite{Leichenauer:2018obf, Khandker:2018xls, Balakrishnan:2019gxl}) with an extra gravitational term, when $\theta = 0$:
\bne 2\pi \langle T_{vv} \rangle_{\psi} = \frac{\hbar S''}{\cA} + 4G_N \left (\frac{\hbar S'}{\cA} \right)^2 .\ene
In short, there is a speculative but intriguing possibility, hinted at by the SNEC through the Jacobi condition, that the equality relation between the stress tensor and entropy derivatives seen in QNEC saturation generalises to semiclassical gravity.

\acknowledgments

We would like to thank Ben Craps, Victor Franken, Ben Freivogel, Matt Headrick, Aidan Herderschee, Gabriele Pascuzzi, and Aron Wall for useful discussions. 
The research of JRF is supported by FNRS MISU grant 40024018 ``Pushing horizons in Black hole Physics''. The work of AR is supported by FWO-Vlaanderen project G012222N, the VUB Research Council through the Strategic Research Program High-Energy Physics, and FWO-Vlaanderen through a Senior Postdoctoral Fellowship 1223125N.

\paragraph{LLM usage.} This paper contains no LLM-generated content. All ideas originated from the authors, and all results were derived by them. An LLM was used to check results, and mistakes found were verified and corrected by the authors.

\paragraph{Author contributions.} Project conceptualisation, key results, lead on writing (AR); further development of ideas and results, checking results, writing and editing shared between authors.

\bibliography{EnergyBounds}
\bibliographystyle{JHEP}
\end{document}